**A study on the growth mechanism and the process parameters controlling aluminium oxide thin films deposition by pulsed pressure MOCVD**

Hari Murthy, S.S Miya, Susan Krumdieck

**Abstract**

Aluminum oxide thin films were deposited on silicon substrates under different deposition conditions using pulse pressure metal organic chemical vapour deposition (PP-MOCVD). The current study investigates into the growth mechanism of the deposited film and the control of the film morphology by varying the processing parameters of PP-MOCVD - choice of solvent, concentration, and presence of a shield. Aluminum sec-butoxide (ASB) was used as the aluminum source while hexane and toluene were used as the solvents. The films were deposited at 475$^{o}$C at different precursor concentrations. It was observed that the choice of solvent has no effect on the surface morphology, but it influenced the deposition rate. The improved deposition rate, relatively close enthalpy of vaporisation (ΔH) values and uniformity of the film, irrespective of the growth conditions, showed that hexane was a better solvent for ASB than toluene. A hybrid mode of vapour deposition and vapour condensation model for thin film growth is proposed where five different mechanisms lead to a solid film formation. These include vapour phase deposition under low arrival rate, vapour phase deposition under high arrival rate, Leidenfrost aerosol formation, heterogeneous particle formation and liquid droplet impingement. The important parameter that needs to be controlled is the precursor flux arrival rate which can be controlled by varying the precursor concentration, use of a solvent with a low ΔH and the presence of a shield over the substrate, which influences the surface morphology and the growth rate of the films.

## 1.1 Introduction

### *1.1.1 Materials*

Alumina in the form of powder or coating is a material that has wide applications in optoelectronic and microelectronic components due to its high band gap (9 eV) as well as wear resistant agent or protection against corrosion and temperature oxidation [1]. Alumina films have been produced by various techniques including sol-gel synthesis [2,3], atomic layer deposition (ALD) [4], electron beam deposition (E-beam) [5], sputtering [6], pulsed laser deposition (PLD) [7], ultrasonic spray pyrolysis [8–11], spray deposition [12] and chemical vapour deposition (CVD) [13].

According to Marchand et al. [14], the morphology of the films deposited depends on several factors, such as temperature, solvent, and the substrate material. Precursor-solvent chemistry plays a vital role when it comes to direct liquid injection (DLI) deposition techniques, with a preference for solvents with high vapour pressure [15]. Pulse pressure MOCVD (PP-MOCVD) is a unique variant of direct liquid injection CVD (DLI-CVD) where the precursors are dissolved in a suitable solvent and fed into a continuously evacuated chamber via an ultrasonic atomiser [16]. No carrier gas is used to transport the solution to the deposition chamber. The exposure of the liquid droplets to a vacuum causes flash vaporisation [17,18] resulting in a near-instantaneous pressure spike followed by a pump down for each pulse. The rapid change in the chamber pressure results in the existence of an expansion mass transport regime [19]. The injection has to be done instantaneously for the droplets to be exposed to the minimum chamber pressure [20] which would otherwise lead to agglomeration of particles (aerosols).

In PP-MOCVD, the deposition mechanism is determined by the precursor/solvent compatibility, concentration, and the deposition chamber base pressure [20,21]. The precursor/solvent compatibility can be indicated by the difference in their evaporation rate estimated by comparing their enthalpies of vaporization (ΔH) [21]. ΔH is the amount of heat that needs to be absorbed for a liquid to be vaporised at a given temperature and pressure. A liquid with a low ΔH requires less heat to be vaporised completely than another liquid with a higher ΔH value. A significant difference in the evaporation rate of the precursor and the solvent causes one of the components to evaporate, leaving the other component behind to freeze during deposition. If the solvent evaporates faster than the precursor, a dry, solid spherical particle containing the precursors (aerosols) is left behind leading to aerosol deposition [22]. High liquid-vapour conversion can be achieved if a precursor-solvent pair is selected such that both evaporate at the same rate. Previous work [23] on alumina precursors has shown aluminium sec-butoxide (ASB) to be compatible with hexane and toluene with the best growth rate observed at 475$^{o}$C.

*1.1.2 Condensation model*

The formation of thin films from vapours is a complex process consisting of nucleation, growth, coalescence and sometimes crystallisation [24]. Condensation and nucleation from the vapour on a solid surface have been widely studied and used for thin film formation [25]. The condensation of the vapour occurs when its supersaturation degree reaches a critical value, leading to the subsequent nucleation [26] and formation of liquid droplets [27]. According to the supersaturation-condensation-fusion deposition mechanism [28], the deposition is completed in 2 stages – the first stage follows conventional CVD and is restrained to a thin layer. As the deposition progresses, the supersaturation of the chemical system increases, resulting in the formation of liquid droplets, which fuse to form the thin film. Vapour condensation can be categorised as either drop-wise or film-wise condensation [29]. In the drop-wise condensation, the formation mechanism can be explained by the liquid film rupture hypothesis and the hypothesis of specific nucleation sites [30]. In the drop-wise condensation, a stable cluster forms from the diffusion of adatoms on the surface or from the addition of the atoms from the vapour phase [31]. In the later, the nuclei formation occurs in specific sites, such as pits, grooves or caves. As the arrival rate increases, the thin film forms rapidly, contracts and ruptures into nuclei and droplets, which is in line with the "hypothesis of film rupture" [32]. The surface tension of the growing film ruptures the film ruptures into droplets upon reaching a critical thickness. Because the adatoms of the amorphous film on the surface are highly mobile, we can assume the surface to have a liquid-like fluidity [33]. There are 2 types of island coalescence – solid-like (occurs during the final stages of a mono-layer formation), and liquid-like (initial and intermediate stages of condensation of amorphous and crystalline films). It has been modelled that for a liquid-like coalescence, under conditions of slow growth corresponds to the diffusion mechanism of island growth under the condition of complete condensation. For a high growth rate, disc-shaped clusters are formed on the surface. The vacuum deposition of amorphous condensates has already been presented as 2-D Van der Waals condensation, where the amorphous layers are formed initially followed by the crystalline condensates produced within the amorphous film.

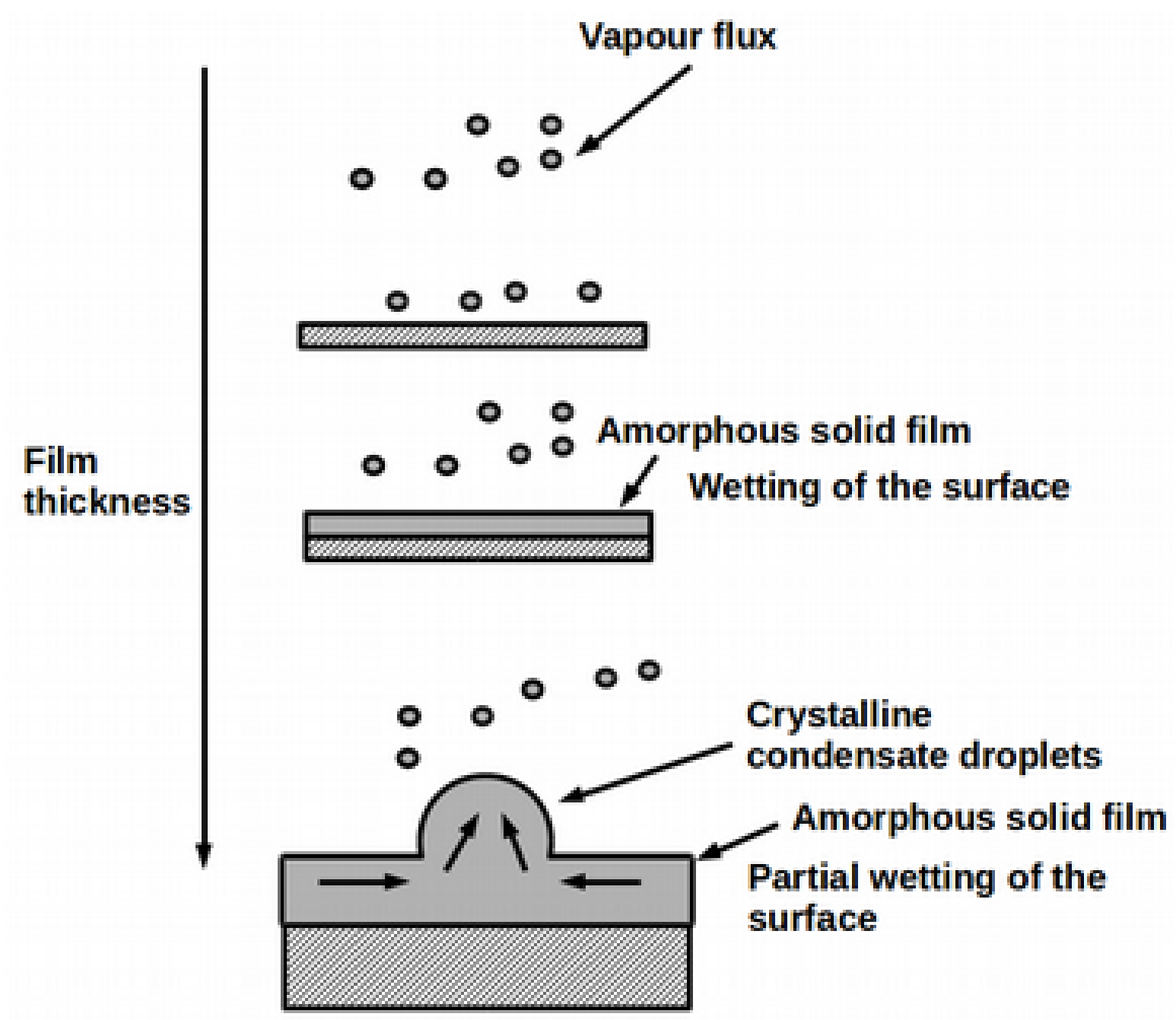

Fig 1: Condensation model of thin film deposition causing the formation of droplets on a surface

This study seeks to make an attempt to understand the film growth mechanism from a droplet-like deposition process and the factors that control it. A hybrid vapour deposition- vapour condensation growth model has been proposed for the alumina thin films prepared by PP-MOCVD, where the arrival rate of the precursor flux to the substrate determines the surface morphology. The experimental variables include the choice of solvents (hexane and toluene), concentration and the presence/absence of a shield. The purpose of the shield as demonstrated by Lee et al. [20] is to provide secondary evaporation of the particles that formed the aerosols, while at the same time prevent the aerosols from contributing to the film growth and morphology. The surface morphology, film composition and the deposition rate of films deposited under different experimental conditions are analysed. The effect of the variables on the morphology and growth rate of the alumina films are analysed. A numerical model describing the droplet behaviour of titanium isopropoxide (TTIP) particles in hexane/toluene, developed by Boichot and Krumdieck [21] is also included with ASB substituted for TTIP to corroborate the experimental observations.

## 1.2 Materials and methods

A research-scale vertical reactor (described in [23]) was used to deposit alumina films on (100) silicon substrates at 475°C. The temperature is so chosen as the main purpose is to understand the growth mechanism for pulsed CVD. Pre-deposition, the substrates were cleaned using a piranha solution (3:1 $H_2SO_4$: $H_2O_2$) followed by a dilute HF dip. ASB (Sigma-Aldrich, >97% pure, CAS Number 2269-22-9) was used without any further purification, with dry hexane and toluene used as the solvent.

Two concentrations (0.125 mol% and 0.5 mol%) of the precursor-solvent solution were selected, prepared in a glovebox filled with dry $N_2$ gas (purity >99.99%) obtained from BOC Ltd to avoid exposure to air and moisture. The precursor solution was maintained in a pressurised bottle (75-85 kPa) under constant stirring. The pulsing time was set at 6 sec and the reaction chamber base pressure maintained at 100-110 Pa. The details of the varied experimental parameters are given in Table 1. Fig 1 shows the schematics of the deposition process and the research equipment used. A shield is placed at a distance of 15 mm from the top of the susceptor to understand the growth mechanism.

**Table 1**: Details of experimental variables during deposition.

| Solvent | Concentration | Pulses | |
|---|---|---|---|
| | | With shield | No shield |
| Hexane | 0.125 mol% | 1000 | 475 |
| | 0.5 mol% | 500 | 500 |
| Toluene | 0.125 mol% | 1050 | 500 |
| | 0.5 mol% | 500 | 500 |

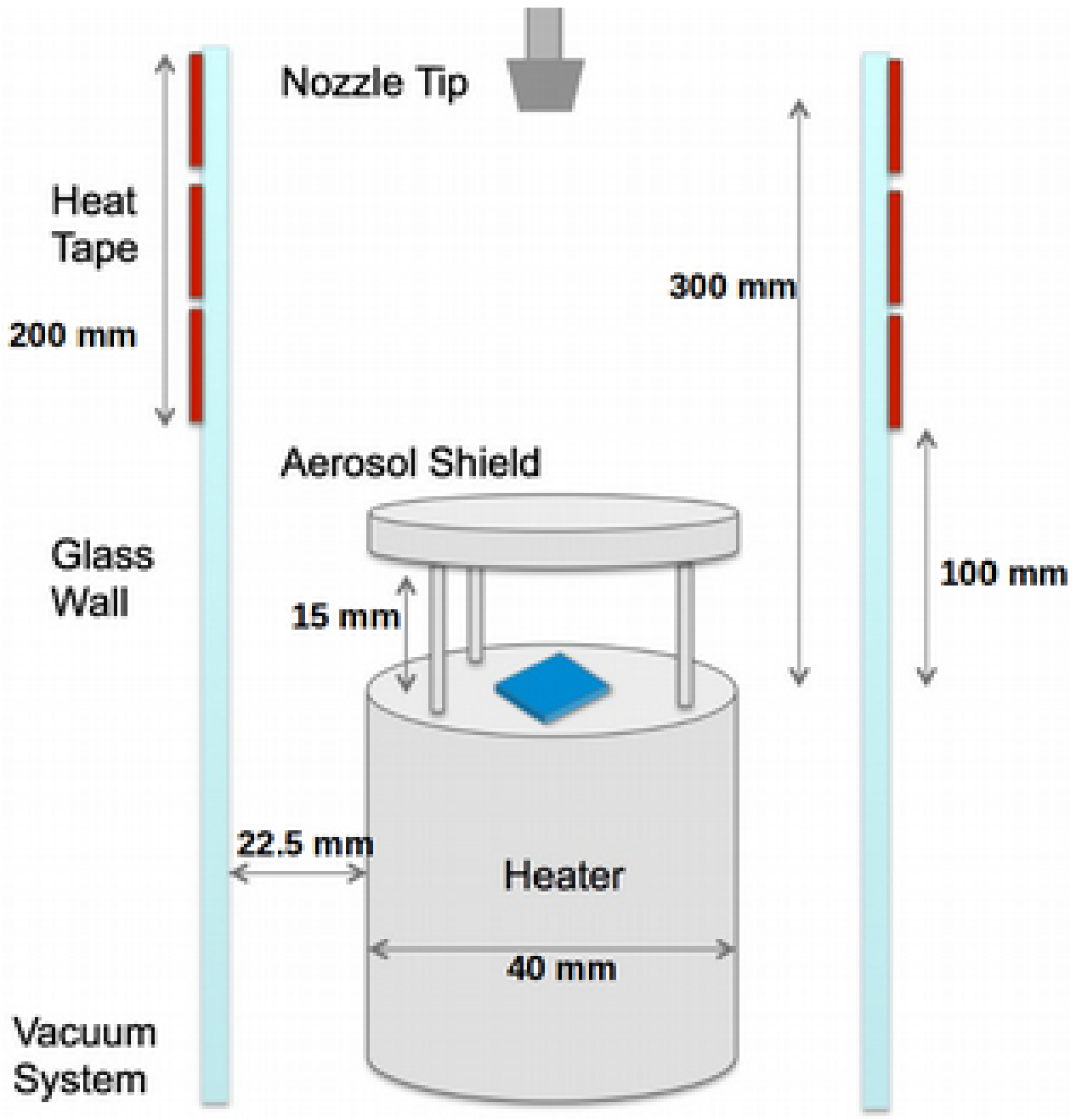

Fig 2:  Schematics of the pulsed MOCVD (PPMOCVD) deposition process used in this study.

No post-deposition annealing of the samples was done. The surface morphology, composition, and thickness were measured using a JEOL JSM 7000F field emission, high resolution scanning electron microscope (FE-SEM) fitted with Gatan cathodoluminescence detector and JEOL energy dispersive X-ray analysis system (EDX). The film thickness was determined using cross-sectional SEM and taking the average. The accelerating voltage was maintained at 5 keV.  All samples were pre-coated with gold for 120 sec at 25 mA to prevent any charging effects. The FTIR spectrum of the thin films was obtained from Bruker Tensor II ATR-FTIR spectrophotometer using absorbance measurement accessory. EDX and FTIR were performed at multiple locations on the sample, and the values were averaged. The surface roughness was measured using a NanoScope IIIa scanning probe microscope with a scan rate. Crystal structure was determined using XRD ($\lambda$= 1.54184 Å) with a $10^{o}$ incident angle.

## 1.3 Results and Discussion

### *1.3.1 Modeling droplet diameter*

A model developed by Boichot and Krumdieck [21] is used to model the precursor-solvent droplet behaviour, regarding droplet diameter/mass and chamber pressure variations. Previous models using TTIP in hexane and toluene has suggested that a low ΔH/Cp and a low ΔH of the components results in better vaporisation of the droplet. The model considers 3 concentrations of the precursor – 0 (solvent only), 0.125 mol% and 0.5 mol%, with a 6 sec pulse cycle. Two reactor wall temperatures ($T_{reactor}$ =23°C and 223°C) are tested to see the influence of the external heat on the chamber pressure ($P_{reactor}$). Table 2 lists the key input properties of the precursor and solvents used in the model.

The decrease in the droplet diameter ($m_{droplet}$) is given by:

$$\frac{dm_{droplet}}{dt} = -\phi_{vap,x}$$

$$\phi_{vap,x} = A_{droplet} \frac{2}{2-\alpha_c\sqrt{2\pi M_x R}} \left(\alpha_E P_{Sat} \frac{T_{droplet}}{\sqrt{T_{droplet}}} - \alpha_c \frac{P_{reactor}}{\sqrt{T_{reactor}}}\right) M_x \quad \text{--(2)}$$

$$m_{droplet} = m_{solvent} + m_{precursor}$$

'*x*' represents the component of the droplet measured – either the pure solvent or the precursor-solvent solution. $\varphi_{vap,x}$ is the mass vaporisation flux, $A_{droplet}$ is the mole fraction of the solvent/precursor, $M_x$ is the molar mass of the species, *R* is the universal gas constant, and $P_{sat}$ is the vapour pressure of the solvent/precursor. The values of condensation coefficient ($\alpha_C$) and evaporation coefficient ($\alpha_E$) are taken about 0.1 for a liquid/solid surface vaporising in a vacuum. The negative sign indicates the decrease in the mass of the droplet with time.

**Table 2**: Input variables used in the model for droplet evaporation.

| **Property** | **ASB** | **Hexane** | **Toluene** |
|---|---|---|---|
| ΔH (J/kg) | 365000* | 334000 | 402000 |
| Melting Point (in $^{o}$C) | -46 | -93.5 | -95 |
| $C_p$ (J/kg K) | 2096** | 2293 | 1720 |

** Specific heat capacity ($C_p$) of triethylaluminium is used (obtained from NIST, CAS# 97-93-8)

* Obtained from Gelest Inc.

Fig 3 shows the reduction in the droplet diameter over the course of a pulse for a pure solvent and precursor solution. The process can be divided into 3 stages, which affects the growth mechanism of the thin film – instantaneous flash vaporisation, pseudo-equilibrium/solvent evaporation (0-2.6 sec for 0.125 mol% ASB-toluene solution) and precursor evaporation (2.6-3.4 sec for 0.125 mol% ASB-toluene solution). The flash vaporisation is the first stage where the droplet undergoes a liquid to gas phase conversion as it is exposed to the low pressure in the chamber [34]. A droplet with a low ΔH ensures that more of the droplet gets converted to the vapour phase, as seen with hexane. At the beginning of the pulse, the droplets are exposed to the minimum pressure in the chamber and undergo flash vaporisation. The decrease in the droplet diameter is influenced by the ΔH of the solvents. Hexane, due to its lower ΔH, has a faster decrease in the droplet diameter (23% reduction) compared to toluene (10% reduction). Following the flash vaporisation, the solvent starts evaporating where hexane evaporates at the rate of ~12 μm/sec and toluene evaporates at ~5 μm/sec. Solvents with a lower ΔH value takes less time to evaporate compared to one with a higher ΔH value [23].

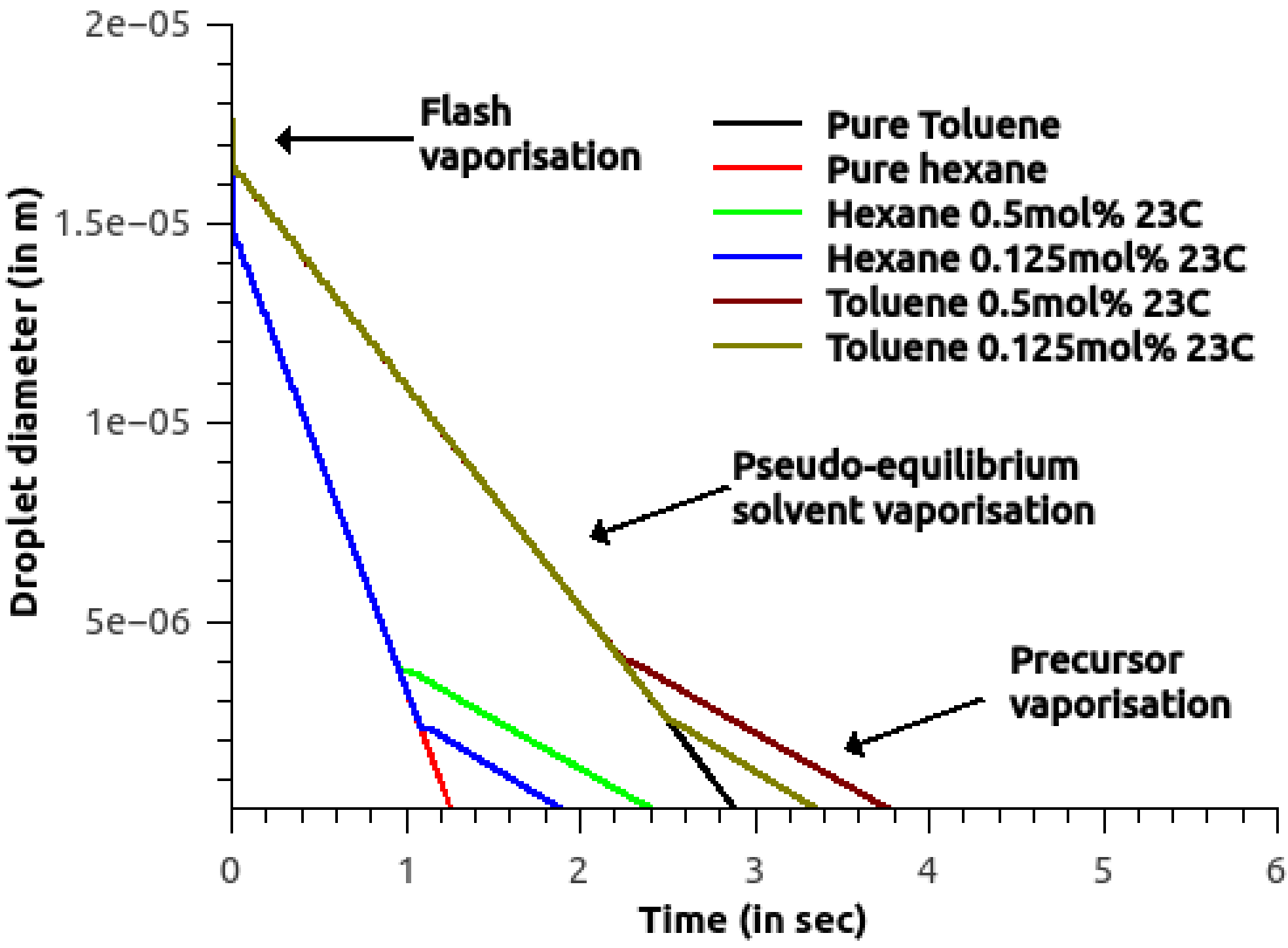

Fig 3: Droplet diameter v/s time for aluminum sec butoxide in hexane and toluene at different concentrations. Solutions with decreased concentration would have a longer solvent vaporisation time before the precursors start evaporating.

Finally, once all the solvent has evaporated, the precursor starts evaporating. The higher ΔH for toluene causes a slower evaporation of the ASB-toluene solution, resulting in a higher possibility of formation of aerosols. The model also predicts a higher chamber pressure using hexane due to its higher volatility and vapour pressure which is validated from the experimental data ($P_{max,hexane}$ = 600 Pa, $P_{max,toluene}$ = 400- 450 Pa). The concentration of the precursors is quite low to affect the properties of the solution, and we can assume the solution to have similar properties to that of a pure solvent.

It is evident from the model that hexane is a better solvent for ASB compared to toluene, due to its low ΔH. It is hypothesised that for any precursor-solvent to be efficient in a pulsed MOCVD process, the $\Delta H_{solvent}/\ \Delta H_{precursor}$ must be as low as possible and as close to each other as possible. Other factors that control the deposition mechanism are the precursor concentration and the presence of a shield.

*1.3.2 Film properties*

A uniform film has been obtained using the deposition process with the surface roughness ranging from 5-7 nm, with a smoother film when a shield is used. Fig 4 shows the AFM micrograph using hexane at 0.5 mol% indicating a uniform film, typical for alumina thin films deposited at low temperature [35,36].

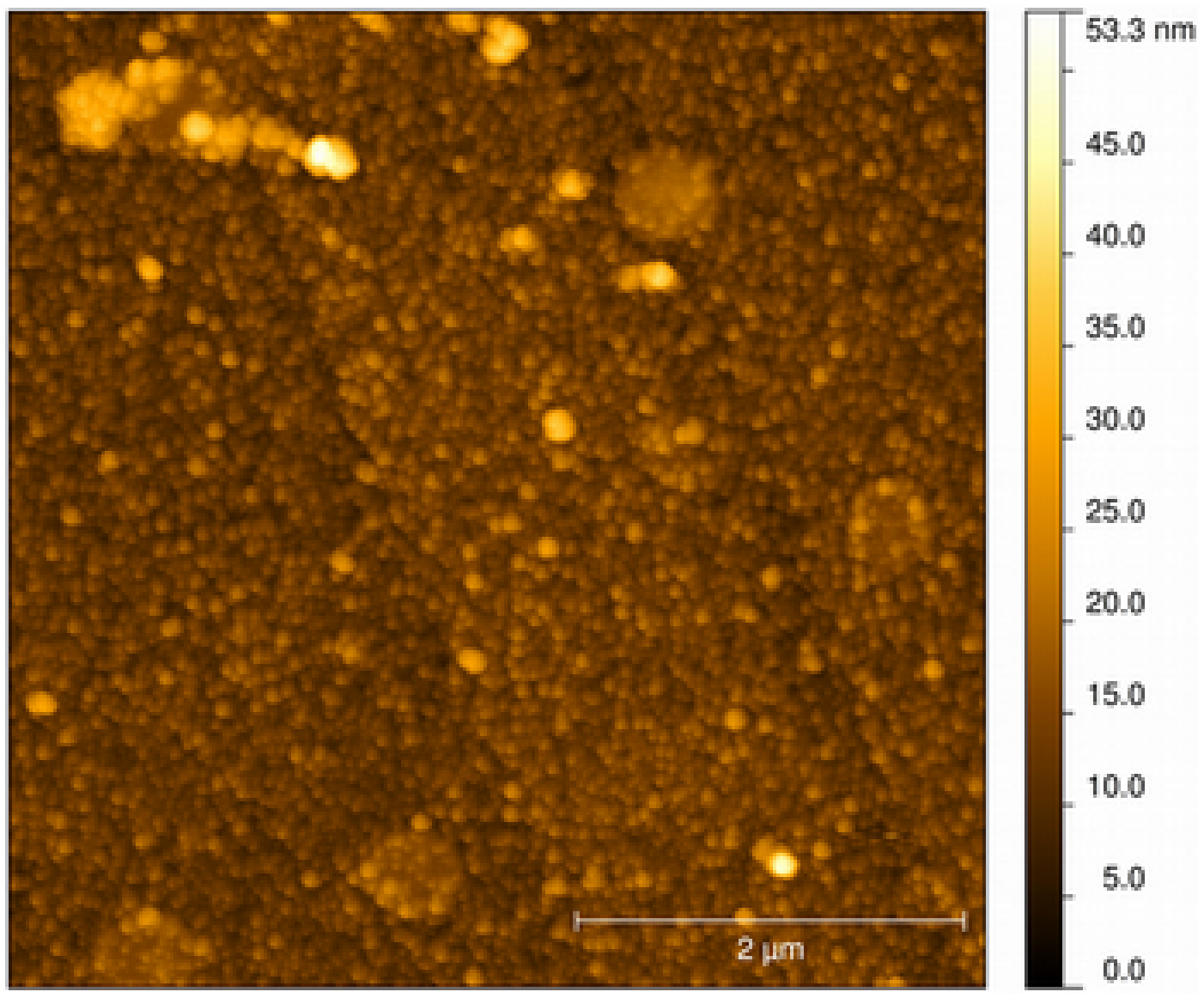

Fig 4: AFM micrograph for uniform thin film using hexane at 0.5 mol% when a shield is placed over the substrate, resulting in thin film deposition from the vapour phase.

The O: Al ratio of the thin films obtained from EDX is normalised with a standard alumina sample analysed under the same operating conditions [37], given by, R (=$(O/Al)_{sample}$/ $(O/Al)_{reference}$). The films are stoichiometric if R=1, oxygen-rich if R>1 and aluminium-rich if R<1. The as-deposited films are found to be oxygen-rich (R>1). This is explained by the β-hydride elimination mechanism for the gas phase decomposition of ASB, which results in excess oxygen getting deposited on the growing film [38]. The films have low carbon content (~2-6 at.%) with hexane showing a higher carbon content than toluene. This is attributed to its lower activation energy compared to toluene ($E_a$ hexane= 230-260 kJ/mole [39,40], $E_a$ toluene

= 293-335 kJ/mole [41–43]). This implies that hexane has a higher possibility to decompose and deposit carbon in the growing film, particularly at the temperature used during the deposition process (450-500ºC). A part of the carbon may also come from the carbonyl ligands of the precursor. The absence of any oxidants, such as water vapour, also assists in the incorporation of the carbon in the growing film as there are no avenues for the removal of the carbon from the system [44]. At the given deposition temperature, the carbon obtained are expected to be aliphatic carbons where the carbon is attached to another carbon or hydrogen bonds, which is as expected from the gas phase decomposition of ASB [38,45]. The presence of the shield has no effect on the carbon content of the film.

The FTIR of the depositions under different deposition conditions is shown in Fig 5. Peaks observed at ~736 $cm^{-1}$ correspond to asymmetric Al-O in $AlO_6$ octahedra [46,47] and those seen at ~880 $cm^{-1}$ correspond to $AlO_4$ bonds [48]. The peaks between 750-850 $cm^{-1}$ are from the Al-O stretching mode [49]. The peak at ~1100-1050 $cm^{-1}$ is associated with the Si-O stretching mode [8]. A shift in the bond vibration from 890 $cm^{-1}$ to 868 $cm^{-1}$ is observed when the depositions are carried out at without the shield. One of the possible causes for the shift is the increase in the size of the grains [50], while the shift can also be due to the compressive stress acting on the film [51] during the deposition. No O-H peaks are observed in the FTIR spectra.

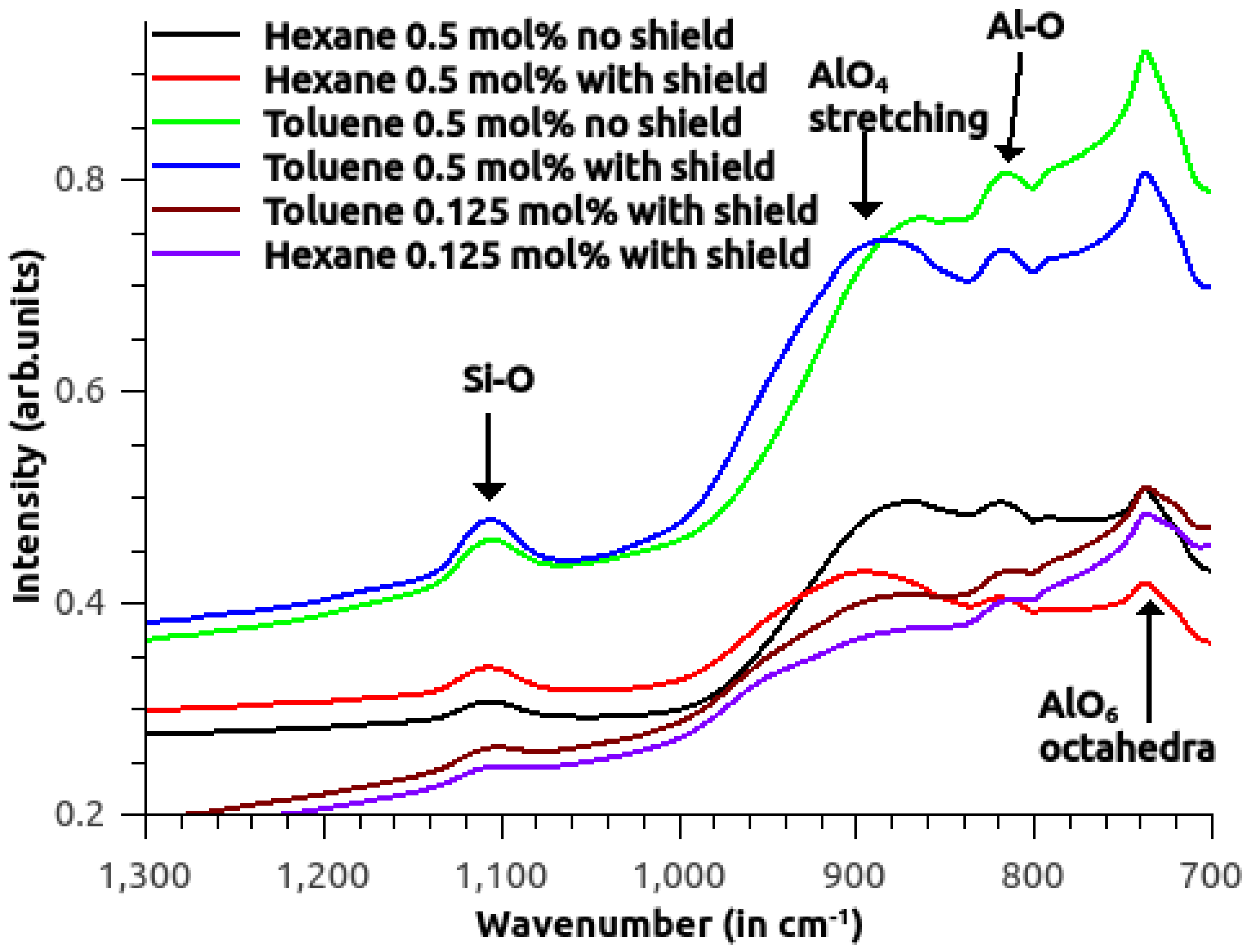

Fig 5: FTIR spectrum from depositions using (a) toluene, and (b) hexane at different concentrations and experimental conditions.

The XRD for the films deposited at 0.5 mol% under different experimental conditions is shown in Fig 6. The graphs are smoothened using a Savitzky-Golan filter of the 1st order to remove the noise without distorting the peaks. The crystallite peaks of the deposited film are associated to k-alumina (JCPDS # 52-0803), a metastable phase with an orthorhombic crystal structure [52]. The peaks are of very low intensity with significant background noise. Favaro et al. [47] have reported on the interpretation of the XRD by comparing IR measurements. The peaks found in the IR spectrum at 736 $cm^{-1}$ and 820 $cm^{-1}$, are related to the $AlO_6$ and $AlO_4$ vibration modes, which are observed in the kappa- phase of alumina [47]. This proves that the aluminium oxide formed is of the kappa phase.

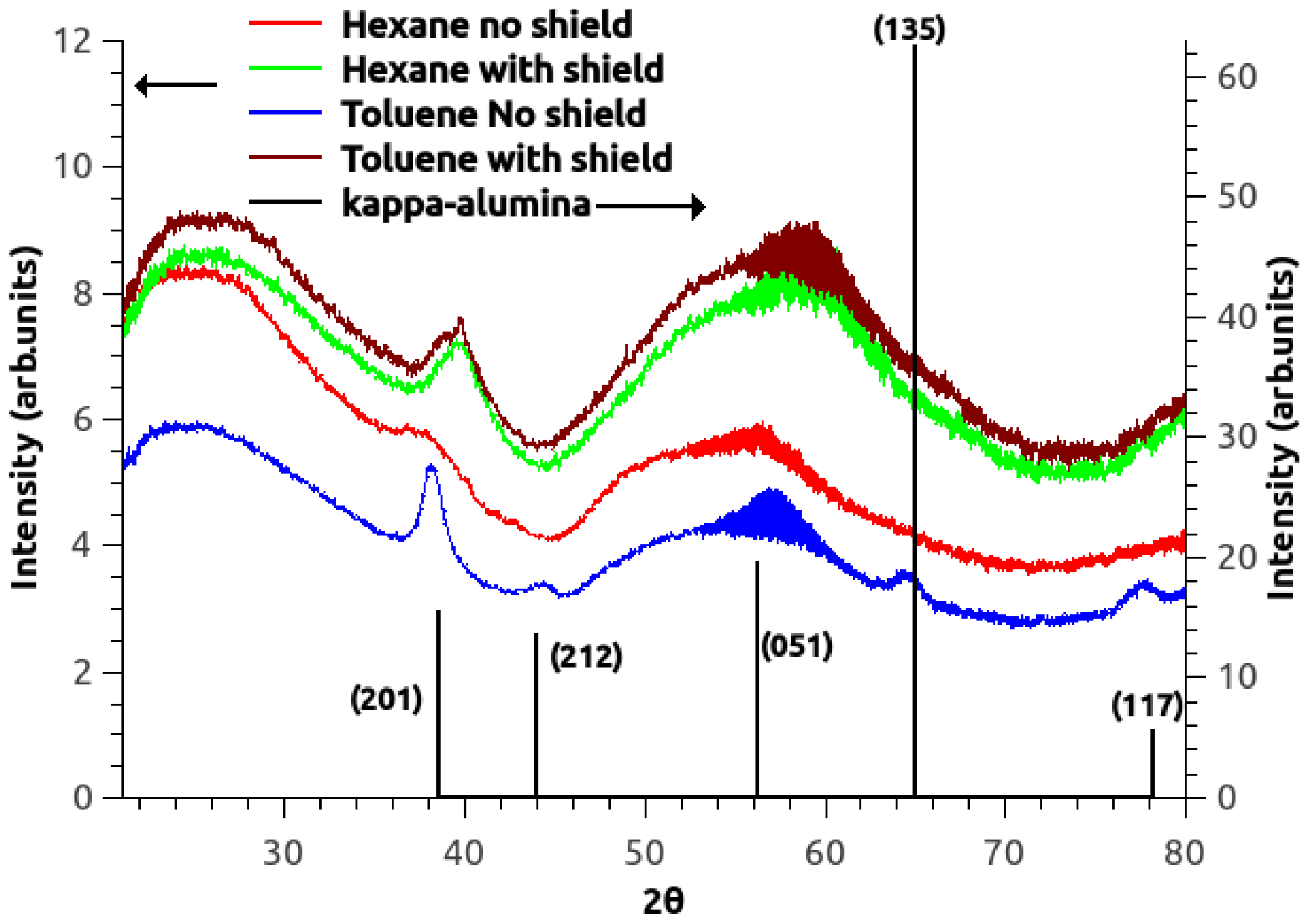

Fig 6: XRD of the films deposited under different experimental conditions shows the presence of nano/micro-crystallites of k-$Al_2O_3$

Solutions with lower ΔH ensure more material getting vaporised during the flash vaporisation, as shown in Fig 7, which reflects the effect of the solvents, concentration and the shield on the growth rate of the deposited film. The presence of the shield leads to a 23% and 69% reduction in the growth rate using hexane and toluene as solvents, respectively. This supports the hypothesis that hexane is a better medium in converting more of the injected droplets to vapour compared to toluene, primarily due to its lower ΔH, resulting in a greater reduction in the droplet diameter during the flash vaporisation stage. In the absence of the shield, toluene exhibited a higher growth rate compared to hexane, while increasing the concentration results in higher precursor input for forming the films, thereby increasing the growth rate [16].

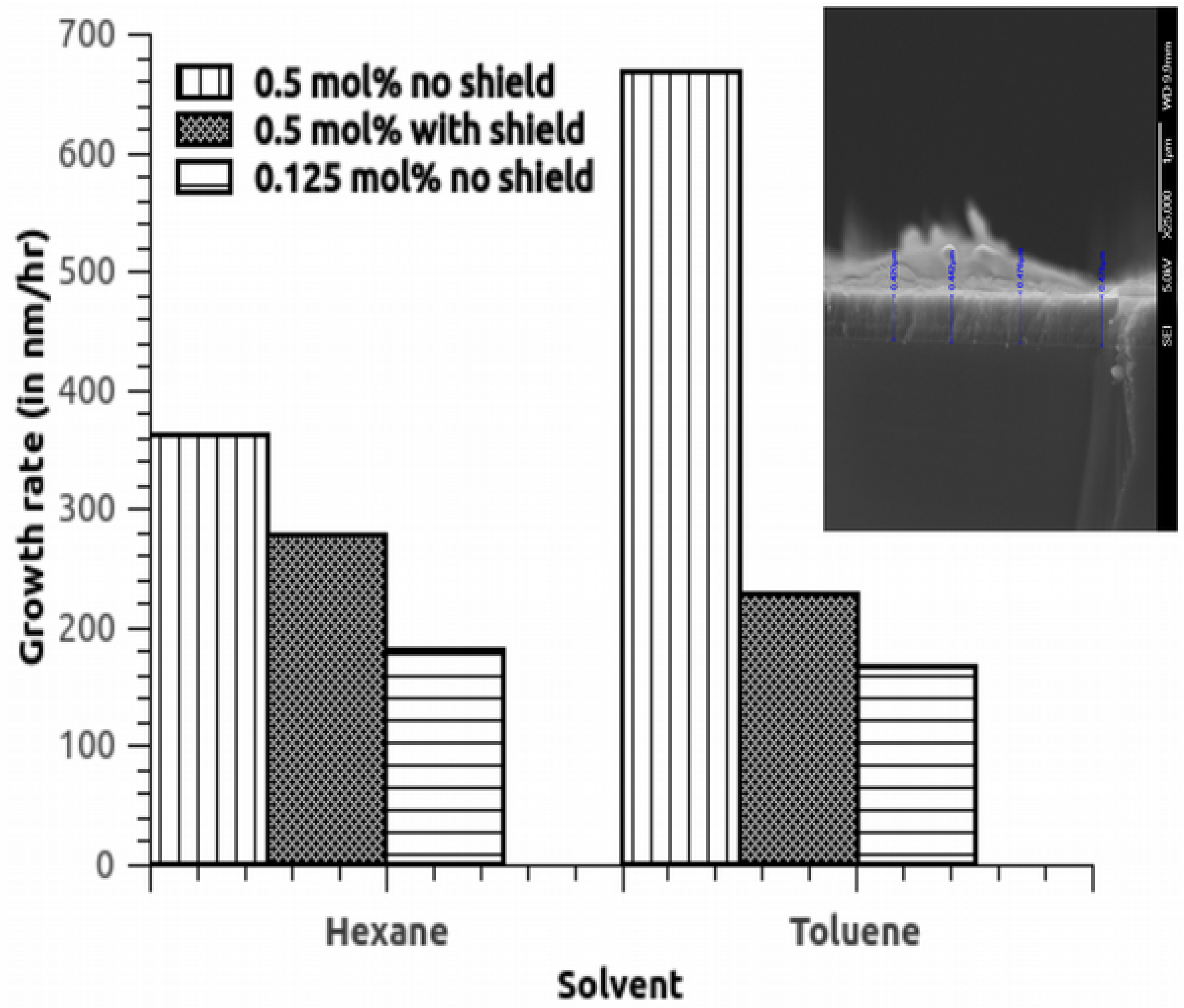

Fig 7: Growth rate for hexane and toluene under different conditions and concentrations. Inset: Cross-sectional SEM micrograph for film deposited using toluene without shield at 0.5 mol%.

*1.3.3 Proposed growth mechanism*

The growth mechanism proposed in PP-MOCVD is a combination of either one or more of the processes, given below (Fig 8):

a) Vapour phase deposition under low arrival rate,

b) Vapour deposition under high arrival rate,

c) Particle formation,

d) Liquid droplet impingement, and

e) Leidenfrost aerosol formation.

| | |
|---|---|
| 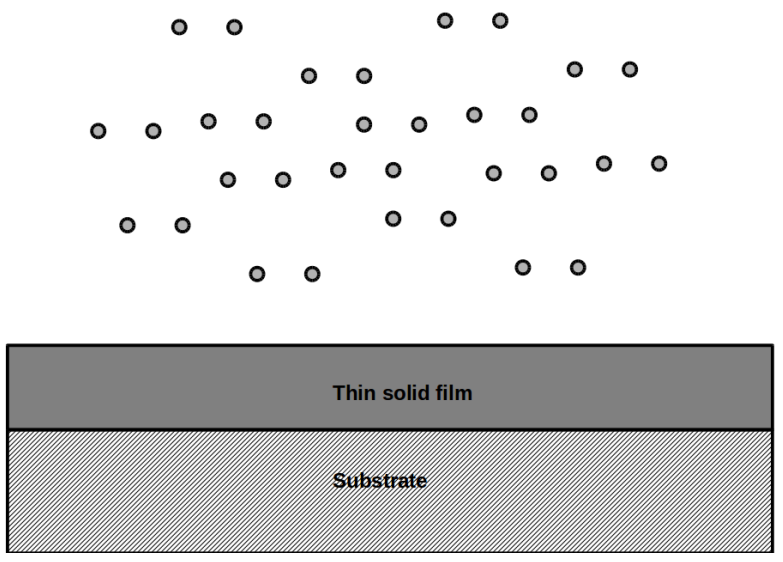 | • Vapour phase deposition.<br>• Uniform arrival rate.<br>• Low growth rate.<br>• Low arrival rate ($\Phi_{vap,1}$).<br>• Factors controlling process: low precursor concentration – no shield, high precursor concentration – with shield.<br>• *Desirable*. |
| 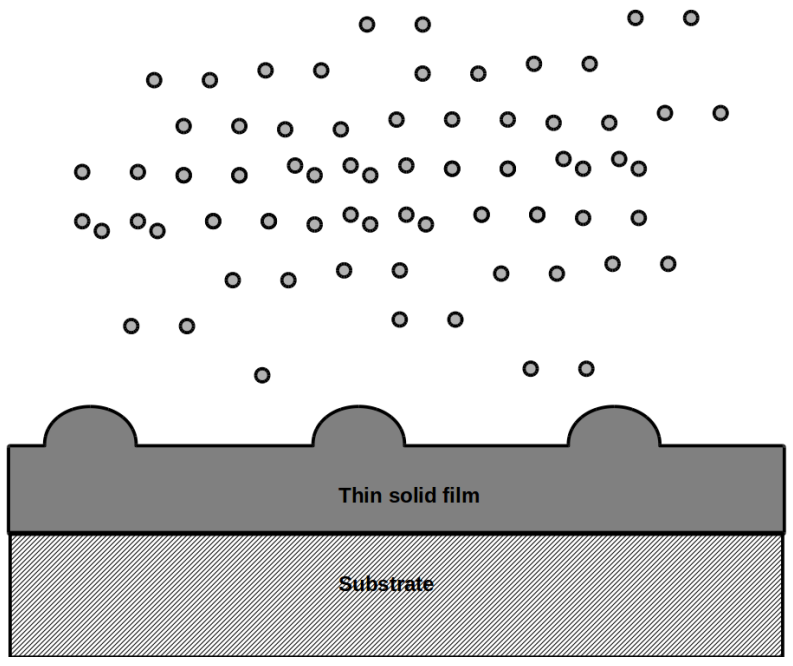 | • Vapour phase deposition.<br>• High arrival rate ($\Phi_{vap,2}$).<br>• Uniform arrival rate.<br>• ($\Phi_{vap,2}$) > ($\Phi_{vap,1}$).<br>• High growth rate.<br>• Factors controlling process – high concentration – no shield.<br>• *Desirable*. |
| 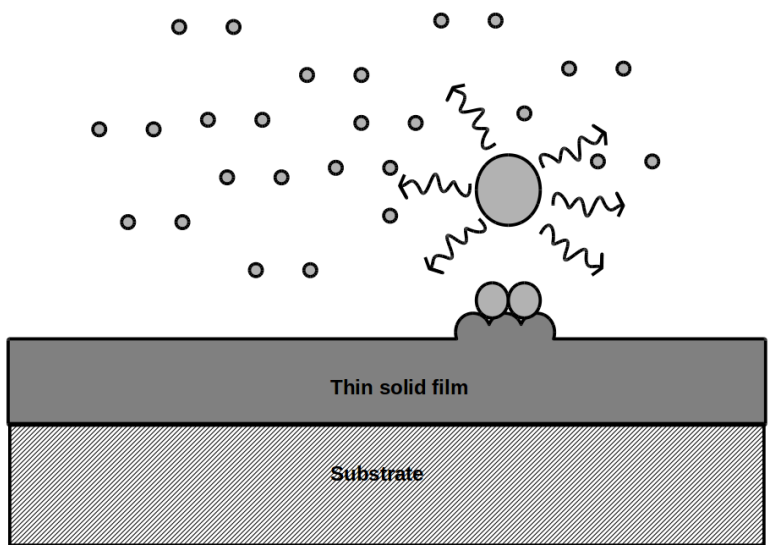 | • Leidenfrost aerosol formation.<br>• Localised high $\Phi$.<br>• Factors controlling process- solvent with high enthalpy of vaporisation, high concentration.<br>• *Desirable under controlled conditions.* |
| 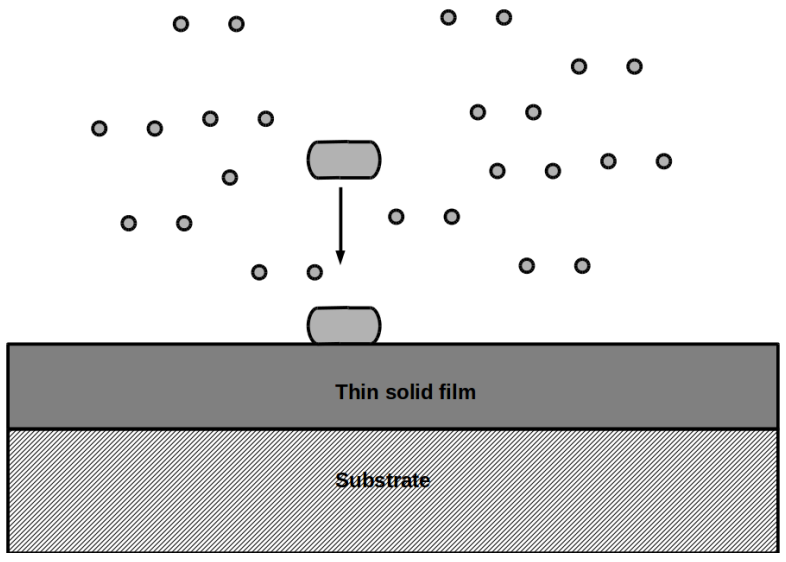 | • Heterogeneous powder formation.<br>• Formed due to high heat flux from the surrounding.<br>• Formation, impingement along with vapour deposition.<br>• Factors controlling process – external heat from surrounding.<br>• *Undesirable*. |
| 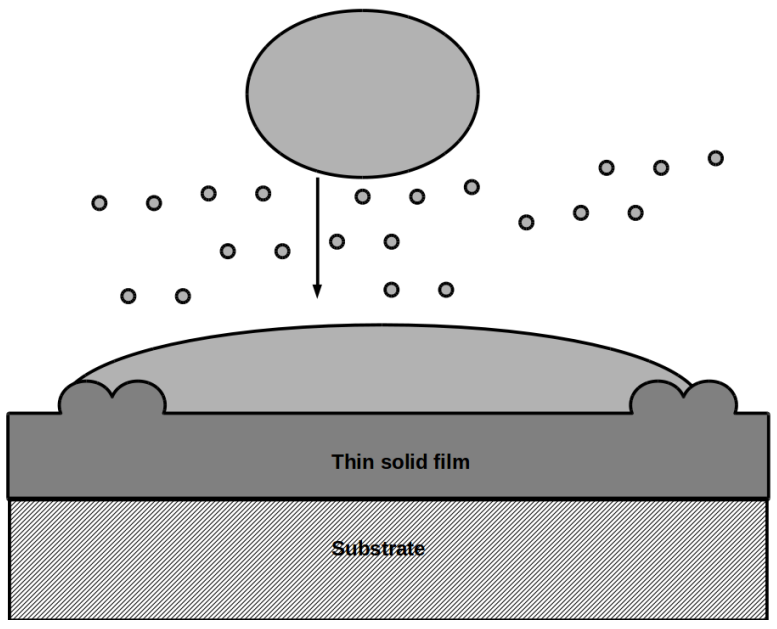 | • Droplet impingement.<br>• Splating of the droplets on the surface.<br>• Factors controlling process – malfunction of ultrasonic nozzle.<br>• *Undesirable*. |

Fig 8: Proposed growth mechanism for thin film deposition using pulse pressure MOCVD. The process parameters – properties of solvent, concentration and presence of shield can be varied to control the growth mechanism and film morphology.

The combined effect of all the processes is the formation of a solid thin film with the surface morphology and growth mechanism governed by the choice of solvent, precursor concentration and the presence of a shield over the substrate. Liquid-gas phase conversion during the flash vaporisation of the precursor solution results in the vapour deposition of a solid thin film. The arrival rate of the precursor vapour to the substrate affects the surface tension of the material, which in turn, determines the morphology of the film [53]. A lower arrival rate results in the growth of a smooth uniform film, achieved by using a low concentration (Fig 9a) or by placing a shield in the path of the precursor vapour (Fig 9b). The deposited grains migrate slowly to stable positions rather than agglomerating [54].  In the absence of a shield, the substrate is exposed to a high precursor flux, and high growth rate is observed (Fig 7). This increases the cohesive forces acting on the film, and it ruptures forming drop-like condensates on the surface (Fig 9c,9d).

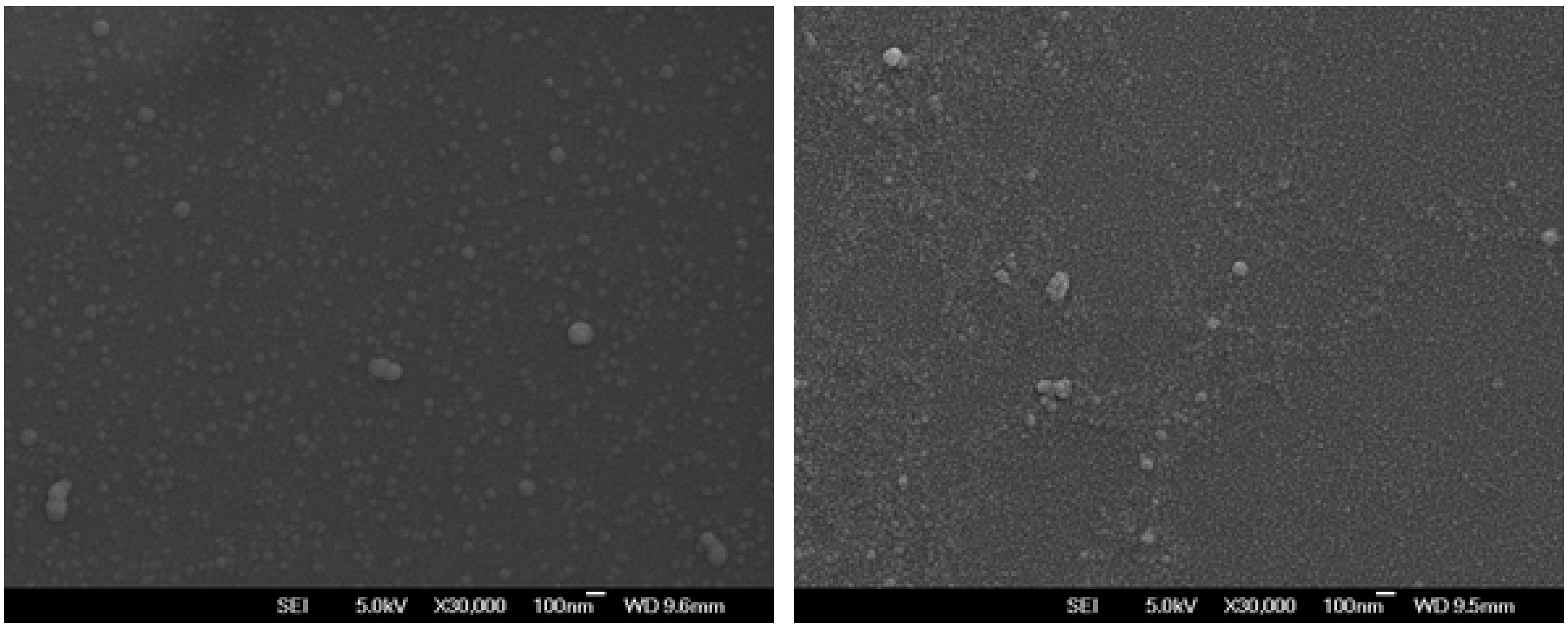

(a) 0.125 mol% Hexane no shield

(b) 0.5 mol% Hexane with shield

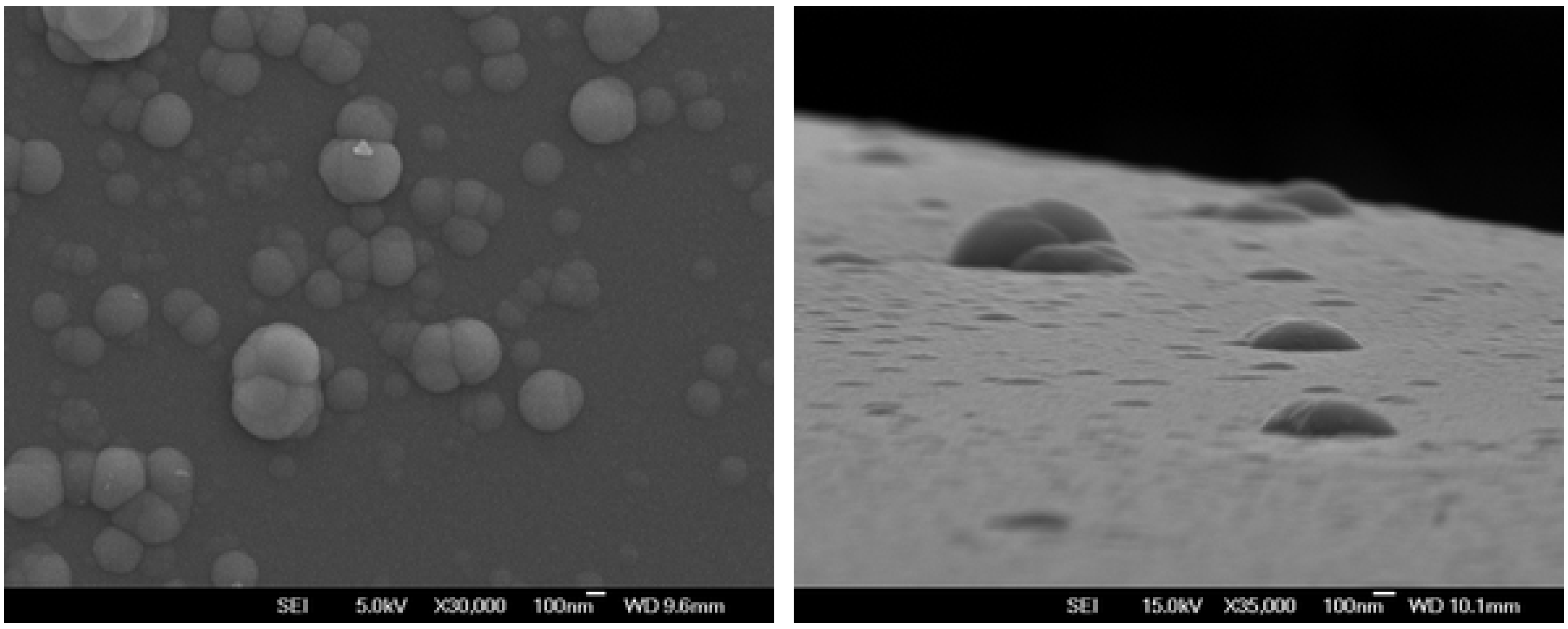

(c) 0.5 mol% Hexane no shield (d) 0.5 mol% Hexane no shield

Fig 9: Effect of concentration and shield on the film morphology using hexane as the solvent.

In the absence of the shield, the film growth happens from a combination of vapour phase deposition coupled with other phenomena, and can be controlled using a shield over the substrate. As the droplets approach the heated substrate, it experiences an increase in the ambient temperature due to the radiation heat from the heater. The solvent starts evaporating close to the surface, resulting in a high local flux of dried precursors which have been coagulated at the core of the droplet as the solvent was evaporated [55]. The irregular solvent evaporation leaves behind concentrated precursor droplets [56], which, along with the Leidenfrost effect, causes the spheroidal features on the surface similar to the features commonly reported for aerosol-based deposition [12,57] (Fig 10).

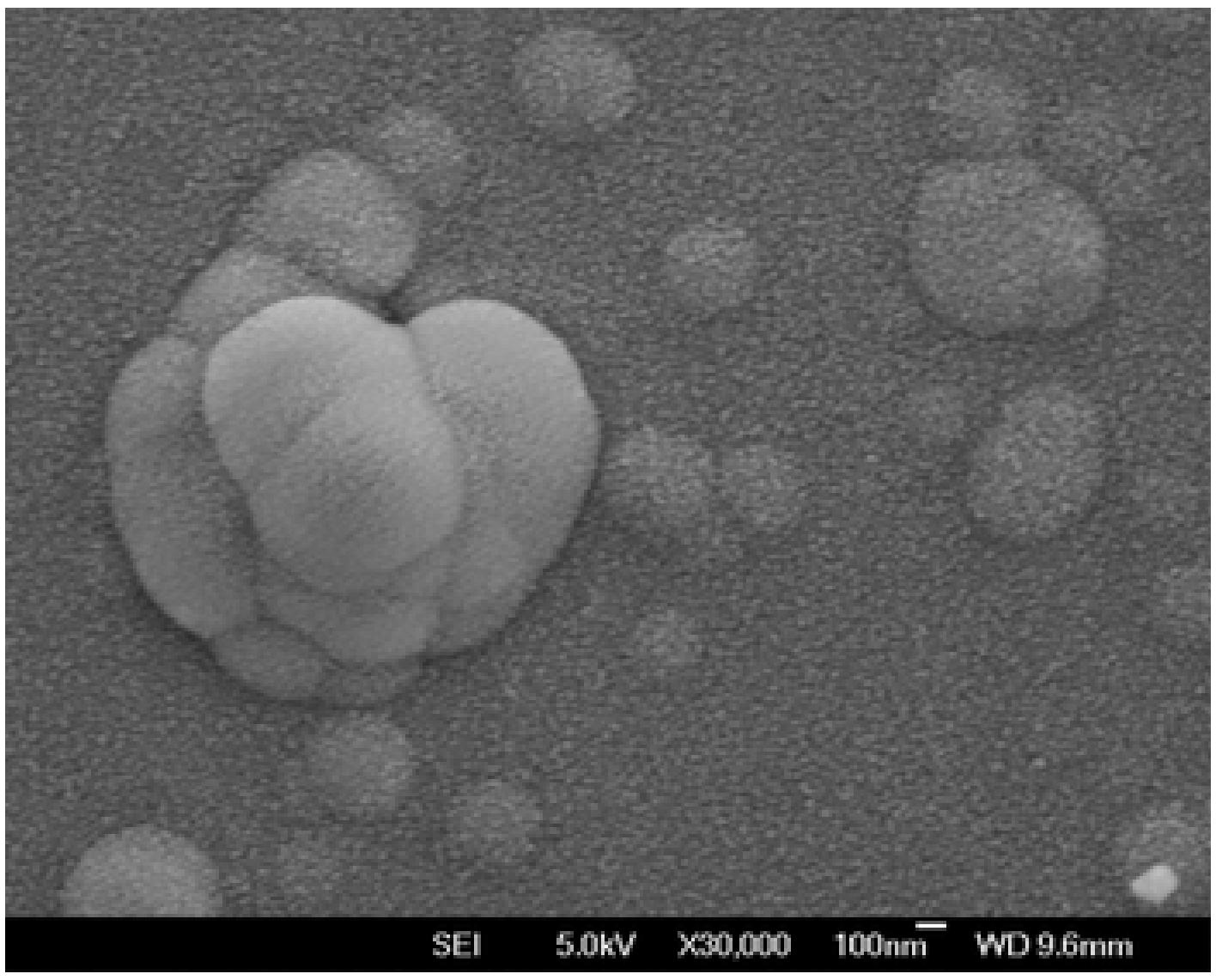

Fig 10: Leidenfrost aerosol assisted deposition due to a high localised flux of the precursor aerosols, using hexane 0.5 mol% no shield.

Another possibility is the evaporation of the solvent much before reaching the substrate due to heat from the ambience, which decomposes the precursors as it approaches the substrate. These form weakly-adherent powder particles on the substrate [58] and are undesirable for any practical applications. The density of such decomposed particles getting onto the film (Fig 11a-11b), suggests that hexane is a better solvent for vapour phase deposition.

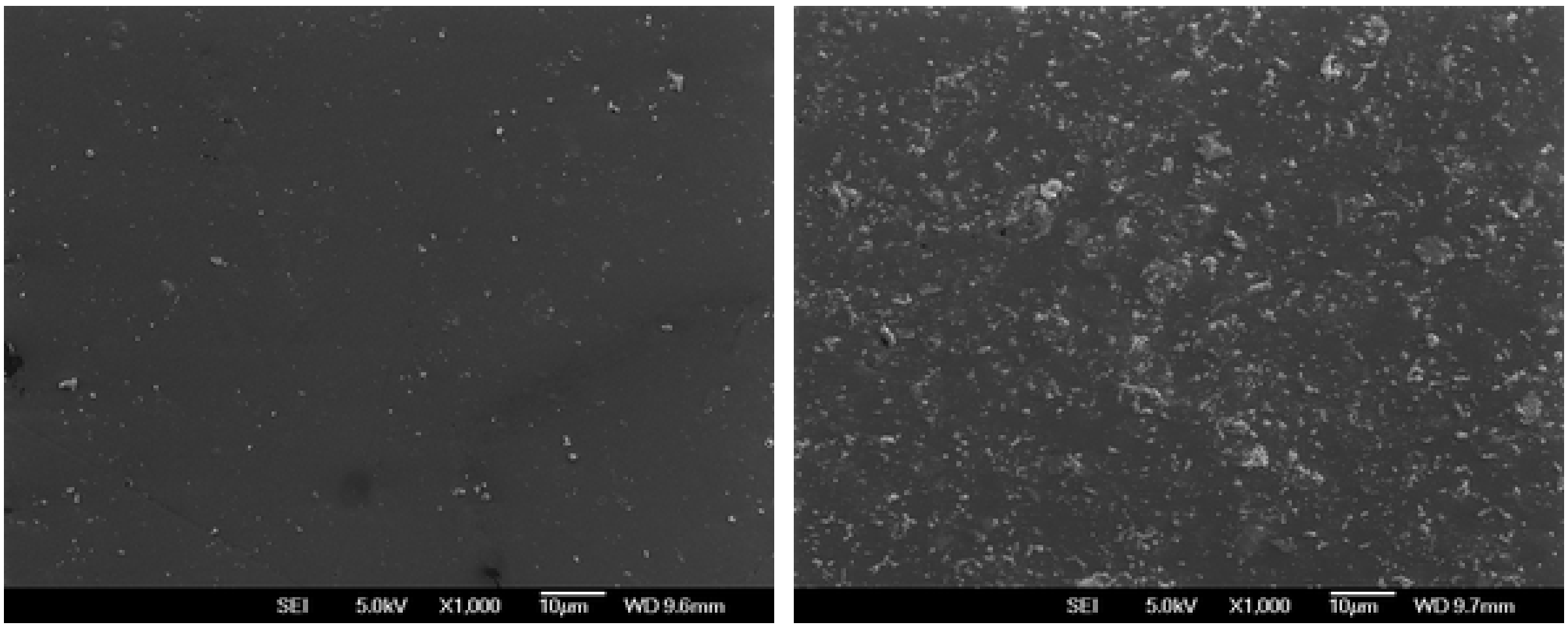

(a) Hexane 0.5 mol% no shield

(b) Toluene 0.5 mol% no shield

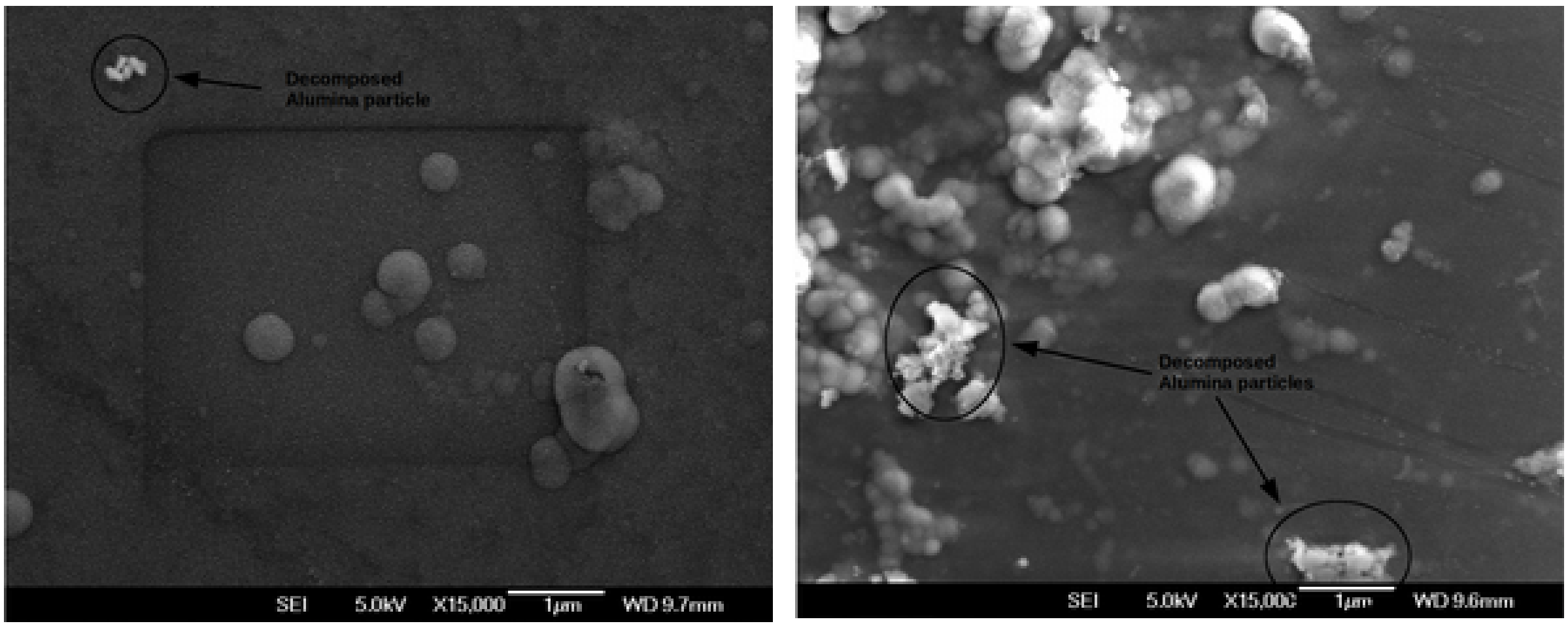

(c) Hexane 0.5 mol% no shield (d) Toluene 0.5 mol% no shield

Fig 11: Effect of the solvent on the deposition mechanism at the same concentration without the shield, showing the decomposed particles on the surface of the substrate.

There is also the possibility of a liquid droplet to survive the flash vaporisation and subsequent solvent evaporation and arrive at the surface intact, where it undergoes solvent evaporation and decomposition of the deposited material, in a process similar to direct droplet impingement/spray pyrolysis [59]. At temperatures above the boiling point of the droplet, the droplet spreads on the surface, and vapour bubble nucleation occurs within the liquid [59]. The droplet boils and evaporates while the particles are deposited in a circular pattern (Fig 12).

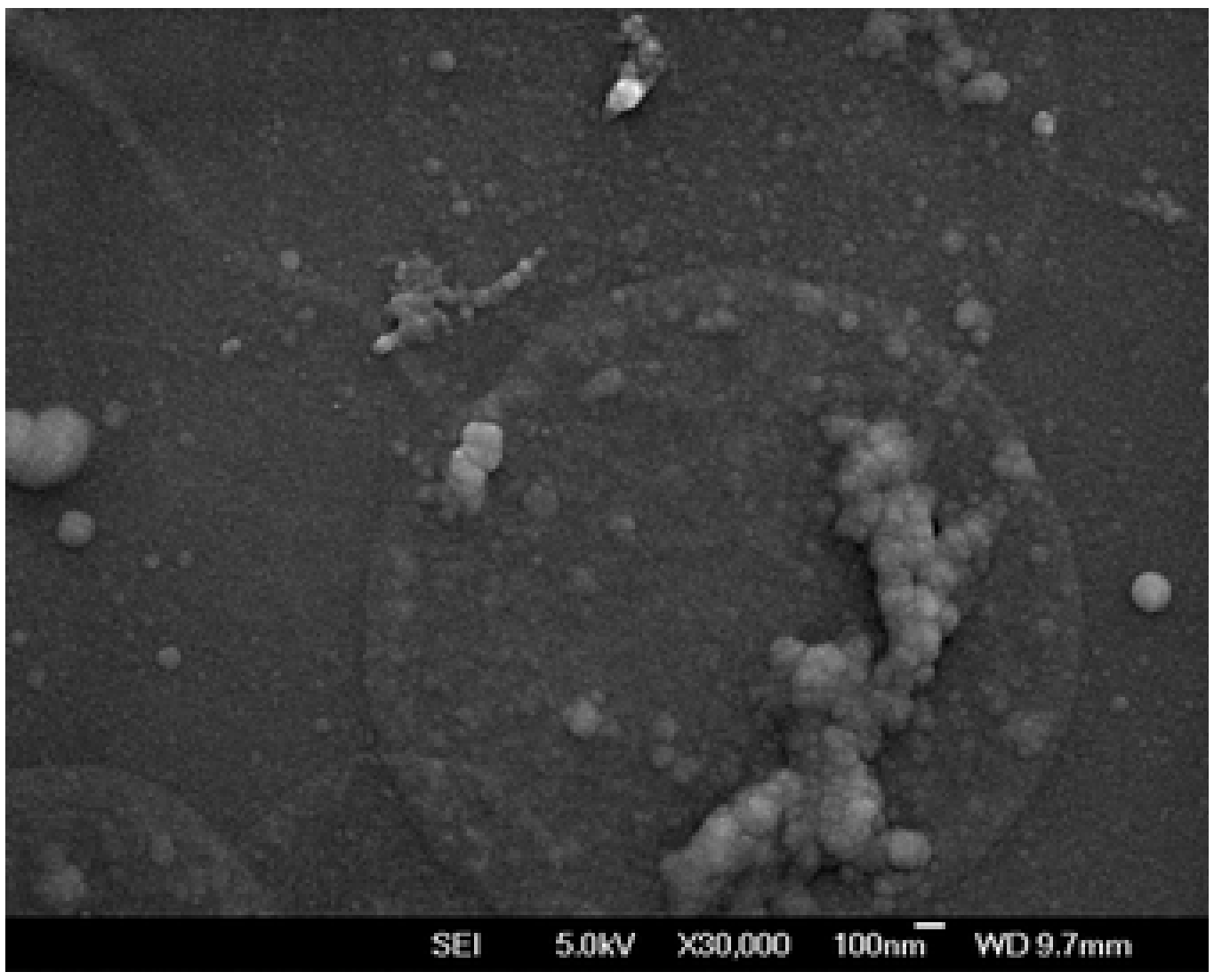

Fig 12: Direct liquid impingement causing the formation of circular pattern deposition on the substrate, using hexane 0.125 mol% no shield.

## Conclusions

A hybrid vapour deposition-vapour condensation film growth mechanism in pulsed pressure MOCVD is proposed using alumina case study. Aluminium sec-butoxide is used as the aluminium source dissolved in hexane and toluene. Hexane, due to its lower enthalpy of vaporisation, vaporises faster than toluene, suggesting that it is a better solvent for pulsed MOCVD. Three stages of droplet evaporation have been modelled – flash vaporisation, solvent evaporation, and precursor evaporation, which determines the film growth mechanism. A smooth, uniform film with an RMS roughness of 5-7 nm is obtained, with films deposited using a shield being more smoother than those deposited without the shield. The deposited films are oxygen-rich, and FTIR confirms the presence of $AlO_6$ octahedra and $AlO_4$ bonds on the surface. Low-intensity XRD peaks of the thin films reveal micro-crystallites of k-phase of alumina, confirmed by FTIR spectrum. Films deposited using hexane have a higher growth rate under the shield than toluene, which confirms that it is a better solvent for vapour phase deposition. It has been proposed that the pulsed CVD process has different growth mechanism occurring, with the process controlled by varying the process parameters, such as a solvent with low ΔH, precursor concentration, and by the presence of a shield over the substrate. The different processes leading to the film deposition include vapour phase deposition, Leidenfrost aerosol deposition, heterogeneous particle formation and liquid droplet impingement. The arrival rate of the vapour flux to the substrate plays a crucial role in determining the surface morphology during the vapour phase deposition. The arrival rate can be controlled by using a low precursor concentration or by placing a shield over the substrate. Under conditions of low arrival rate, the grains diffuse across the substrate slowly without forming condensates, while for a high arrival rate, the film's cohesive force coupled with its increased growth rate causes the film to rupture forming droplet-like features on the surface. Liedenfrost aerosol formation results from the evaporation of the solvent just above the substrate resulting in a high localised vapour flux. External heat from the surroundings cause the particles to decompose in the gas phase and land on the surface without contributing to the film growth. Circular patterns are also found on the substrate due to the liquid impingement on the substrate leaving the particles on the surface which gets decomposed while the solvent gets evaporated. Vapour phase deposition and the Liedenfrost aerosol deposition are the desired growth mechanisms, while the others processes are undesirable.

## References

[1] V.A.C. Haanappel, H.D. Van Corbach, T. Fransen, P.J. Gellings, Corrosion resistant coatings ( A1203 ) produced by metal organic chemical vapour deposition using aluminium-tri-sec-butoxide, Thin Solid Films. 230 (1993) 138–144.

[2] J. Masalski, J. Gluszek, J. Zabrzeski, K. Nitsch, P. Gluszek, Improvement in corrosion resistance of the 316l stainless steel by means of Al 2 O 3 coatings deposited by the sol-gel method, Thin Solid Films. 349 (1999) 186–190.

[3] P. Vitanov, A. Harizanova, T. Ivanova, T. Dimitrova, Chemical deposition of Al2O3 thin films on Si substrates, Thin Solid Films. 517 (2009) 6327–6330. doi:10.1016/j.tsf.2009.02.085.

[4] J. Koo, S. Kim, S. Jeon, H. Jeon, Y. Kim, Characteristics of Al2O3 Thin Films Deposited Using Dimethylaluminum Isopropoxide and Trimethylaluminum Precursors by the Plasma-Enhanced Atomic-Layer Deposition Method, 48 (2006) 131–136.

[5] a. K. Chin, Al2O3 as an antireflection coating for InP/InGaAsP LEDs, J. Vac. Sci. Technol. B Microelectron. Nanom. Struct. 1 (1983) 72. doi:10.1116/1.582507.

[6] J. Echigoya, H. Enoki, RF magnetron sputtered aluminium oxide coatings on iridium, J. Mater. Sci. 31 (1996) 5247–5256.

[7] J. Gottmann, A. Husmann, T. Klotzbiicher, E.W. Kreutz, Optical properties of alumina and zirconia thin films grown by pulsed laser deposition, Eur. Phys. Journal. Appl. Phys. 101 (1998) 0–4. doi:10.1051/epjap/2013120530.

[8] S. Carmona-Tellez, J. Guzman-Mendoza, M. Aguilar-Frutis, G. Alarcon-Flores, M. Garcia-Hipolito, M. a. Canseco, C. Falcony, Electrical, optical, and structural characteristics of Al2O3 thin films prepared by pulsed ultrasonic sprayed pyrolysis, J. Appl. Phys. 103 (2008) 34105. doi:10.1063/1.2838467.

[9] M. Aguilar-frutis, M. Garcia, C. Falcony, G. Plesch, S. Jimenez-sandoval, A study of the dielectric characteristics of aluminum oxide thin films deposited by spray pyrolysis from Al(acac)3, Thin Solid Films. 389 (2001) 200–206.

[10] B.P. Dhonge, T. Mathews, S.T. Sundari, C. Thinaharan, M. Kamruddin, S. Dash, a. K. Tyagi, Spray pyrolytic deposition of transparent aluminum oxide (Al 2O3) films, Appl. Surf. Sci. 258 (2011) 1091–1096. doi:10.1016/j.apsusc.2011.09.040.

[11] J.C. Ortiz, A, Alonso, High quality-low temperature aluminum oxide films deposited by ultrasonic spray pyrolysis, J. Mater. Sci. Mater. Electron. 13 (2002) 7–11.

[12] Y. Wu, K.-L. Choy, The microstructure of alumina coatings prepared by aerosol assisted spray deposition, Surf. Coatings Technol. 180–181 (2004) 436–440. doi:10.1016/j.surfcoat.2003.10.078.

[13] T. Maruyama, S. Arai, Aluminum oxide thin films prepared from aluminum acetylacetonate by chemical vapor deposition, Appl. Phys. Lett. 60 (1992) 322–323.

[14] P. Marchand, I. a Hassan, I.P. Parkin, C.J. Carmalt, Aerosol-assisted delivery of precursors for chemical vapour deposition: expanding the scope of CVD for materials fabrication., Dalton Trans. 42 (2013) 9406–22. doi:10.1039/c3dt50607j.

[15] P.-L. Etchepare, L. Baggetto, H. Vergnes, D. Samélor, D. Sadowski, B. Caussat, C. Vahlas, Process-structure-properties relationship in direct liquid injection chemical vapor deposition of amorphous alumina from aluminum tri-isopropoxide, Phys. Status Solidi. 12 (2015) 944–952. doi:10.1002/pssc.201510037.

[16] S.P. Krumdieck, B.P. Reyngoud, A.D. Barnett, D.J. Clearwater, R.M. Hartshorn, C.M. Bishop, B.P. Redwood, E.L. Palmer, Deposition of Bio-Integration Ceramic Hydroxyapatite by Pulsed-Pressure MOCVD Using a Single Liquid Precursor Solution, Chem. Vap. Depos. 16 (2010) 55–63. doi:10.1002/cvde.200906839.

[17] S.P. Krumdieck, O. Sbaizero, A. Bullert, R. Raj, YSZ layers by pulsed-MOCVD on solid oxide fuel cell electrodes, Surf. Coatings Technol. 167 (2003) 226–233. doi:10.1016/S0257-8972(02)00908-8.

[18] C.W. Lim, H. Cave, M. Jermy, S. Krumdieck, Liquid Droplet Evaporation in Simulations of the Flow in Pulsed-Pressure MOCVD, ECS Trans. (2009) 1251–1258. doi:10.1149/1.3207730.

[19] S. Krumdieck, A. Kristinsdottir, L. Ramirez, M. Lebedev, N. Long, Growth rate , microstructure and conformality as a function of vapor exposure for zirconia thin films by pulsed-pressure MOCVD, Surf. Coatings Technol. 201 (2007) 8908–8913. doi:10.1016/j.surfcoat.2007.03.009.

[20] D. Lee, S. Krumdieck, S.D. Talwar, Scale-up design for industrial development of a PP-MOCVD coating system, Surf. Coatings Technol. 230 (2013) 39–45. doi:10.1016/j.surfcoat.2013.06.064.

[21] R. Boichot, S. Krumdieck, Numerical Modeling of the Droplet Vaporization for Design and Operation of Liquid-pulsed CVD, Chem. Vap. Depos. 21 (2015) 1–10. doi:10.1002/cvde.201507191.

[22] G. V. Jayanthi, S.C. Zhang, G.L. Messing, Modeling of Solid Particle Formation During Solution Aerosol Thermolysis: The Evaporation Stage, Aerosol Sci. Technol. 19 (1993) 478–490. doi:10.1080/02786829308959653.

[23] N.R. Gunby, S. Krumdieck, H. Murthy, S.L. Masters, S.S. Miya, Study of precursor chemistry and solvent systems in pp-MOCVD processing with alumina case study, Phys. Status Solidi. 212 (2013) 1–8. doi:10.1002/pssa.201532309.

[24] A. V. Osipov, Kinetic model of vapour-deposited thin film condensation: stage of liquid-like coalescence, Thin Solid Films. 261 (1995) 173–182. doi:10.1016/S0040-6090(94)06486-5.

[25] L.H. Chen, C.Y. Chen, Y.L. Lee, Nucleation and growth of clusters in the process of vapor deposition, Surf. Sci. 429 (1999) 150–160. doi:10.1016/S0039-6028(99)00360-X.

[26] S. Hashmi, Comprehensive materials processing, Newnes, 2014.

[27] Z. Chen, S. Li, Z. Liu, Morphology and growth mechanism of CVD alumina – silica, Ceram. Int. 31 (2005) 1103–1107. doi:10.1016/j.ceramint.2004.12.003.

[28] Y. Xu, L. Cheng, L. Zhang, W. Zhou, Morphology and growth mechanism of silicon carbide chemical vapor deposited at low temperatures and normal atmosphere, J. Mater. Sci. 4 (1999) 551–555.

[29] Q. Sheng, J. Sun, Q. Wang, W. Wang, H.S. Wang, On the onset of surface condensation: formation and transition mechanisms of condensation mode, Sci. Rep. 6 (2016).

[30] L. Tianqing, M. Chunfeng, S. Xiangyu, X. Songbai, Mechanism Study on Formation of Initial Condensate Droplets, Am. Inst. Chem. Eng. 53 (2007) 1–6. doi:10.1002/aic.

[31] Y.L. Lee, J.R. Maa, Nucleation and growth of condensate clusters on solid surfaces, J. Mater. Sci. 26 (1991) 6068–6072. doi:10.1007/BF01113885.

[32] S.K. Basant Singh Sikarwar , Sameer Khandekar , Smita Agrawal, K. Muralidhar, Dropwise Condensation Studies on Multiple Scales, Heat Transf. Eng. 33 (2012) 301–341. doi:10.1080/01457632.2012.611463.

[33] I. Gutzow, I. Avramov, The mechanism of formation, the structure and the properties of amorphous films, Thin Solid Films. 85 (1981) 203–221.

[34] V. Siriwongrungson, M.M. Alkaisi, S.P. Krumdieck, Step coverage of thin titania films on patterned silicon substrate by pulsed-pressure MOCVD, Surf. Coatings Technol. 201 (2007) 8944–8949. doi:10.1016/j.surfcoat.2007.03.051.

[35] S. Blittersdorf, N. Bahlawane, K. Kohse-Höinghaus, B. Atakan, J. Müller, CVD of Al2O3 Thin Films Using Aluminum Tri-isopropoxide, Chem. Vap. Depos. 9 (2003) 194–198. doi:10.1002/cvde.200306248.

[36] A. Ito, R. Tu, T. Goto, Amorphous-like nanocrystalline gamma-Al2O3 films prepared by MOCVD, Surf. Coatings Technol. 204 (2010) 2170–2174. doi:10.1016/j.surfcoat.2009.11.043.

[37] A.N. Gleizes, C. Vahlas, M.M. Sovar, D. Samélor, M.C. Lafont, CVD-fabricated aluminum oxide coatings from aluminum tri-iso-propoxide: Correlation between

processing conditions and composition, Chem. Vap. Depos. 13 (2007) 23–29. doi:10.1002/cvde.200606532.

[38] H.D. Van Corbach, T. Fransen, P.J. Gellings, The pyrolytic decomposition of aluminium-tri- sec-butoxide during chemical vapour deposition of thin alumina films, Thermochim. Acta. 240 (1994) 67–77.

[39] K.H. Ebert, H.J. Ederer, G. Isbarn, The thermal decomposition of n-hexane, Int. J. Chem. Kinet. 15 (1983) 475–502. doi:10.1002/kin.550150508.

[40] R.. Partington, C.. Danby, The thermal decomposition of n-hexane, J. Chem. Soc. (1948) 2226–2232.

[41] M. Szwarc, The C–H Bond Energy in Toluene and Xylenes, J. Chem. Phys. 128 (2002). doi:10.1063/1.1746794.

[42] R. Sivaramakrishnan, R.S. Tranter, and K. Brezinsky*, High Pressure Pyrolysis of Toluene. 1. Experiments and Modeling of Toluene Decomposition, J. Phys. Chem. A. 110 (2006) 9388–9399. doi:10.1021/jp060820j.

[43] S.J. Price, The pyrolysis of toluene, Can. J. Chem. 10 (1962).

[44] J. Kim, H.A. Marzouk, P.J. Reucroft, J.D. Robertson, C.E. Hamrin, Effect of water vapor on the growth of aluminum oxide films by low pressure chemical vapor deposition, Thin Solid Films. 230 (1993) 156–159.

[45] H. Gong, M.P. Singh, S.A. Shivashankar, T. Shripathi, K. Road, A STUDY OF ALUMINA THIN FILMS GROWN BY LOW PRESSURE MOCVD: XPS AND AES, Int. J. Mod. Phys. B. 16 (2002) 1261–1267.

[46] G. Ji, M. Li, G. Li, G. Gao, H. Zou, S. Gan, X. Xu, Hydrothermal synthesis of hierarchical micron flower-like γ-AlOOH and γ-Al2O3 superstructures from oil shale ash, Powder Technol. 215–216 (2012) 54–58. doi:10.1016/j.powtec.2011.09.005.

[47] G. Sattonnay, Experimental and ab initio infrared study of χ - , κ - and α -aluminas formed from gibbsite, J. Solid State Chem. 183 (2010) 901–908. doi:10.1016/j.jssc.2010.02.010.

[48] G. Sattonnay, J.B. Brubach, P. Berthet, A. Boumaza, L. Favaro, J. Le, A.M. Huntz, P. Roy, R. Te, Transition alumina phases induced by heat treatment of boehmite : An X - ray diffraction and infrared spectroscopy study, J. Solid State Chem. 182 (2009) 1171–1176. doi:10.1016/j.jssc.2009.02.006.

[49] X. Multone, High Vacuum Chemical Vapor Deposition ( HV-CVD ) of Alumina Thin Films, 2009.

[50] S. Kurien, ANALYSIS OF FTIR SPECTRA OF NANOPARTICLES of MgAl2O4, SrAl2O4, and NiAl2O4, 1595 (2005) 64–78. http://mgutheses.in/page/?q=T 1354&search=siby+kurien&page=1&rad=sc.

[51] K. Sowri Babu, A.R. Reddy, C. Sujatha, K.V. Reddy, A.N. Mallika, Synthesis and optical characterization of porous ZnO, J. Adv. Ceram. 2 (2013) 260–265. doi:10.1007/s40145-013-0069-6.

[52] M. Halvarsson, J.E. Trancik, S. Ruppi, The microstructure of CVD k-Al2O3 multilayers separated by thin intermediate TiN or TiC layers, Int. J. Refract. Met. Hard Mater. 24 (2006) 32–38. doi:10.1016/j.ijrmhm.2005.06.007.

[53] D. Vick, L.J. Friedrich, S.K. Dew, M.J. Brett, K. Robbie, M. Seto, T. Smy, Self-shadowing and surface diffusion effects in obliquely deposited thin films, Thin Solid Films. 339 (1999) 88–94. doi:10.1016/S0040-6090(98)01154-7.

[54] X.J. Qin, L. Zhao, G.J. Shao, N. Wang, Influence of solvents on deposition mechanism of Al-doped ZnO films synthesized by cold wall aerosol-assisted chemical vapor deposition, Thin Solid Films. 542 (2013) 144–149. doi:10.1016/j.tsf.2013.07.002.

[55] A. Gurav, T. Kodas, T. Pluym, Y. Xiong, Aerosol Processing of Materials, Aerosol Sci. Technol. 19 (1993) 411–452. doi:10.1080/02786829308959650.

[56] S. Krumdieck, S. Davies, C.M. Bishop, T. Kemmitt, J.V. Kennedy, Al2O3 coatings on stainless steel using pulsed-pressure MOCVD, Surf. Coatings Technol. 230 (2013) 208–212. doi:10.1016/j.surfcoat.2013.06.119.

[57] M.I. Martín, L.S. Gómez, O. Milosevic, M.E. Rabanal, Nanostructured alumina particles synthesized by the Spray Pyrolysis method: microstructural and morphological analyses, Ceram. Int. 36 (2010) 767–772. doi:10.1016/j.ceramint.2009.10.013.

[58] M. Khalid, M. Mujahid, A.N. Khan, R.S. Rawat, I. Salam, K. Mehmood, Plasma Sprayed Alumina Coating on Ti6Al4V Alloy for Orthopaedic Implants: Microstructure and Phase Analysis, Key Eng. Mater. 510–511 (2012) 547–553. doi:10.4028/www.scientific.net/KEM.510-511.547.

[59] U.P. Muecke, G.L. Messing, L.J. Gauckler, The Leidenfrost effect during spray pyrolysis of nickel oxide-gadolinia doped ceria composite thin films, Thin Solid Films. 517 (2009) 1515–1521. doi:10.1016/j.tsf.2008.08.158.